\documentclass[useAMS,usenatbib]{mnras} %MNRAS
\usepackage{graphicx}
\usepackage{amsmath}
\usepackage{bm}
\usepackage{url}
\usepackage{color}
\usepackage{arydshln}
\usepackage{times}

\usepackage{lscape}
\usepackage[section]{placeins}
\usepackage{lineno}
\usepackage{soul}
\begin{document}

%%%%% Front Matter %%%%%%%%%%%%%%%%%%%%%%%%%%%%%%%%%%%%%%%%%%%%%%%%%%%%%%%%%%%%

\title{
$N$-body simulations of gravitational redshifts and other 
 relativistic distortions of galaxy clustering}

\author[Zhu et al.]{Hongyu Zhu$^{1,2}$
\thanks{E-mail:hongyuz@andrew.cmu.edu}
, Shadab Alam$^{1,2,3}$, Rupert A. C. Croft$^{1,2}$, Shirley Ho$^{1,2,3,4}$ \newauthor and Elena Giusarma$^{1,2,3,4}$\\
    $^{1}$ Department of Physics, Carnegie Mellon University, 5000 Forbes Ave., Pittsburgh, PA 15213, USA \\
    $^{2}$ McWilliams Center for Cosmology, Carnegie Mellon University, 5000 Forbes Ave., Pittsburgh, PA 15213, USA \\
    $^{3}$ Institute for Astronomy, University of Edinburgh, Royal Observatory, Blackford Hill, Edinburgh, EH9 3HJ, UK \\
    $^{4}$ Berkeley Center for Cosmological Physics, University of California, Berkeley, CA 94720, USA \\
    $^{5}$ Lawrence Berkeley National Laboratory (LBNL), Physics Division, Berkeley, CA 94720, USA
}
    
\date{\today}
\pagerange{\pageref{firstpage}--\pageref{lastpage}}   \pubyear{2015}
\maketitle
\label{firstpage}

%\label{firstpage}
\begin{abstract}
Large redshift surveys of galaxies and clusters are providing the
first opportunities to search for distortions in the 
observed pattern
of large-scale structure due to such effects as gravitational
redshift.  We focus on non-linear
scales and apply a quasi-Newtonian approach using $N$-body simulations
to predict the small asymmetries in the cross-correlation function
of two galaxy different populations. Following recent work by
Bonvin et al., Zhao and Peacock and Kaiser
on galaxy clusters, we include effects which enter at the same order
as gravitational redshift: the transverse Doppler effect, light cone
effects, relativistic beaming, luminosity distance perturbation and wide-angle effects.
 We find that all these effects cause asymmetries in
the cross-correlation functions. Quantifying these asymmetries,
we find that the total effect is dominated by the gravitational redshift
and luminosity distance perturbation at small and large scales
respectively. By adding additional subresolution modelling
of galaxy structure to the large-scale structure information, we find
that the signal is
significantly increased,  indicating that structure on the smallest scales
is important and should be included. We report on comparison of our
simulation results with measurements from the SDSS/BOSS galaxy
redshift survey in a companion paper.

\end{abstract}


%%%%% Main Text %%%%%%%%%%%%%%%%%%%%%%%%%%%%%%%%%%%%%%%%%%%%%%%%%%%%%%%%%%%%%%%
\section{Introduction}

\label{sec:intro}
% Redshift space distortions & shell estimator
Spectroscopic redshift surveys (e.g., \citealt{colless2003}; \citealt{eisenstein2011}; \citealt{deep2013}; \citealt{gama2015})  have been a major 
observational tool in the quest to
understand the galaxy distribution and the evolution of the large scale
structure of the Universe. According to the cosmological
principle, clustering statistics such as the correlation function
and power spectrum should be intrinsically isotropic. However when measured
using redshift as a proxy for line-of-sight distance, deviations from
isotropy become apparent (\citealt{davis1983}; \citealt{kaiser1987}).
The most important distortions are those due to peculiar velocities,
and measurements of these
(e.g., \citealt{peebles1980}; \citealt{kaiser1987}; \citealt{percival2004}; \citealt{reid2014}; \citealt{Alam2016Testing}) have become 
common and useful probes in cosmological experiments as they can
provide constraints on cosmological parameters and models 
(\citealt{blake2013};
\citealt{samushia2014}). In this paper we make detailed predictions of distortions due to redshifts focusing on relativistic effects and including peculiar velocity effects.

% gravitational redshift
Gravitational redshifts are one of the major predictions of General
Relativity (e.g., \citealt{pound1959}) 
and so can be used to test the theory, and probe structure
in the gravitational potential, even on cosmic scales. The gravitational
redshift has long been considered (e.g., \citealt{greenstein1971}; \citealt{lopresto1991}) a component of the
total observed redshift of galaxies, but only recently has it become
feasible to measure. \cite{cappi1995} made predictions
for the gravitational redshift 
profile of 
galaxies in clusters assuming that the mass distribution followed
analytical profiles (\citealt{de1948} and \citealt{hernquist1990}).
This work showed that the central galaxies in the richest observed galaxies
could be redshifted by up to 50 km/s with respect to the other cluster
members, but this would be challenging to measure on
an individual object basis. \cite{kim2004} investigated the statistical
gravitational redshift profile that would result from stacking galaxy
clusters, using $N$-body simulations of a CDM universe to make predictions.
 These predictions showed that the mean redshift of galaxies
averaged in projected separation from the cluster centre could
be used to make a statistical detection of  gravitational redshifts given
data from large galaxy redshift surveys.

The first observational measurement was
carried out by \cite{wojtak2011} who detected the
gravitational redshift in galaxy clusters using 7800 clusters in data
from the SDSS survey. Other measurements of gravitational 
redshifts at similar significance level were made by \cite{dominguez2012},
\cite{jimeno2015}
and also \cite{sadeh2015}. Several authors have since pointed out that 
gravitational redshifts are not the only cause of a mean redshift profile
of galaxies in clusters. 
 \cite{zhao2013} showed that the transverse
Doppler effect (TD) is also relevant, giving an effect which is
opposite in sign to the gravitational redshift effect. \cite{kaiser2013}
(hereafter K13)
investigated several more relevant effects, including measurements on 
the light cone 
(LC)  and special relativistic
beaming (SRB), showing in a comprehensive study that these also appear
in the prediction for the redshift profile at similar order of
magnitude.

% Rupert & Kaiser
The first measurements of relativistic effects in the 
clustering of galaxies have therefore been made in clusters. The 
clustering of all galaxies however is also sensitive to the
gravitational redshift and other relativistic effects.
The spatial distribution of galaxies in redshift space 
appears squashed on large scale
and elongated at small scales along the axis pointing towards the observer 
 due to peculiar velocities (see \citealt{davis1983}; 
\citealt{kaiser1987}). As a result,
in the 2-point correlation function an oval shape of the contours is
expected on large scales and a ``finger of god'' is observed at small scales along
line-of-sight. However, the correlation function preserves the
symmetry along the line of sight direction. If the cross-correlation 
of two different populations of galaxies is computed, and the two
populations lie on average in regions with different gravitational potentials,
then the clustering statistic will not be symmetric about the plane 
perpendicular to the line of sight. \cite{croft2013} 
(hereafter C13) and \cite{bonvin2014}
have explored this using the cross-correlation function, and the effect
was first pointed out by \cite{mcDonald2009} using the power spectrum
(where it leads to an imaginary term).

The mapping of spatial galaxy positions into observed coordinates causes
distortions in clustering and deviations from
isotropy in a more general fashion than those due to the gravitational
redshift. On large scales, where the density fluctuations are in the linear
regime, perturbation theory (PT) can be used to make predictions for 
clustering statistics
(\citealt{bonvin2014}; \citealt{durrer1994}; \citealt{yoo2009}; \citealt{yoo2012}; \citealt{mcDonald2009}). Such relativistic descriptions of clustering
include up to the second order effects
of peculiar velocities, gravitational redshift and lensing amongst others.
Work by \cite{durrer1994} as well as \cite{bonvin2014} has shown that 
distortions in clustering on linear scales, including 
asymmetry in cross-correlation statistics could be measured at 
the few sigma level given future redshift surveys covering large fractions
of the sky with a high density of galaxies. 

Although relativistic effects such as the gravitational
redshift have been easiest to measure in galaxy clusters, there is 
still potentially information available on non-linear scales
from the clustering of field galaxies and the population of galaxies 
in general. C13 concentrated on smaller scales
than the studies which use PT, showing that
with the additions of a  gravitational redshift term (in an
otherwise Newtonian framework) to a halo model of clustering
(\citealt{cooray2002}) allow predictions to be made of asymmetry in the 
 redshift space cross-correlation function.
Such an
asymmetry was quantified  using from an estimator of the mean
redshift of pairs of galaxies in spherical shells (a ``shell estimator'').
C13 compared the results of the halo model to $N$-body simulations, showing
good agreement, but only included the gravitational redshift term in the 
simulation modelling.

% N-body simulations / P-Gadget
In this paper, inspired by K13, we will use $N$-body simulations
to model the non-linear effects that cause asymmetry in the
galaxy cross-correlation function. We will focus on non-linear scales,
where we expect signal to noise to be larger in  measurements
and where PT is not strictly valid. 
 We measure correlation function distortions and shell estimators
from $N$-body simulations run using the code \textsc{P-Gadget3} (see
\citealt{springel2001}; \citealt{springel2005}; \citealt{khandai2011}), in 
dissipationless mode. We  look
at and understand how gravitational redshift effect, transverse
Doppler (TD) effect, light cone (LC) bias and special relativistic
beaming (SRB)  distort the observed cross-correlation function as 
quantified using  the shell estimator and multipoles. Our predictions 
are compared to observational results from the SDSS/BOSS galaxy
redshift survey (CMASS sample of galaxies)
 in a companion paper (\citealt{Alam2016Measurement}).
We note that \cite{cai2016} recently used $N$-body simulations to make
predictions for the asymmetric redshift profile of galaxy clusters, including
all relativistic effects (except for beaming, which depends in detail on the galaxy sample being modelled, and the wide-angle effects
). Our paper has similarities with that
work, but targets  large-scale structure rather than clusters and includes
an application to predictions for a particular sample.

% Paper plan
Our plan for this paper is as follows. In Sec.~\ref{sec:modeling} we
demonstrate how we model these effects (gravitational redshift,
TD, LC, SRB, LDP, the wide-angle effects and other effects mentioned in \cite{cai2016}) in our simulations. The simulation
results are given in Sec.~\ref{sec:simu_tests},
where we show 2d cross-correlation
functions for different effects individually and together. We then quantify the line-of-sight asymmetries of
the cross-correlation function using the shell estimator. 
In Sec.~\ref{sec:halos} we explore how different subhalo potentials
and resolutions of simulations affect the signal.
In
Sec.~\ref{sec:bias} we show results for the simulated 
sample for which galaxy bias most closely 
matches that of the SDSS/BOSS/CMASS sample of
\cite{Alam2016Measurement} and which is used to
compare to observations. In
Sec.~\ref{sec:discussion} we discuss  our results and conclude.

\section{Modeling clustering asymmetry}
\label{sec:modeling}
We consider two galaxy populations with different mean halo masses similar to  \cite{mcDonald2009}, C13 and \cite{bonvin2014}. As we will be using $N$-body simulations
to make our predictions,
we do this according to the halo mass, with a high mass subset $g1$ 
containing half the objects and a low
mass subset $g2$ the other half. C13 found that the cross-correlation function
and associated asymmetry estimator
of redshift distortions could be computed from a halo model,
via the galaxy-mass cross-correlation function and the $g1$-$g2$ galaxy
cross-correlation function. C13 also made predictions for the estimator
by measuring the asymmetry of the cross-correlation due to gravitational
redshifts from $N$-body simulations. For simplicity, and because
the relevant, largely non-linear scales are small, we use the distant-observer
approximation, assuming that the line-of-sight direction is along
the $z$-axis. We describe below the different effects which
contribute to clustering asymmetry and how we
model them in our $g1-g2$ galaxy cross-correlation function from the
simulations. We largely follow the approaches taken by K13 to
model the different non-linear effects, except that we work with 
$N$-body simulations and we compute the clustering of all galaxies,
not just cluster members.

\subsection{Gravitational redshift}
\label{sub:grav}

Gravitational redshifts are induced by differences
in  gravitational potentials. General
Relativity predicts that the wavelength of a photon increases as
it loses energy leaving a potential well. The observed
gravitational redshift at infinity of a photon with wavelength
$\lambda$ emitted from a gravitational potential $\Phi$ is
$z_\mathrm{g} = {\Delta\lambda}/{\lambda}\approx -{\Phi}/{c^2}$.

The magnitude of $z_\mathrm{g}$ is usually very small. As
pointed out by \cite{cappi1995}, in galaxy clusters,
the differences in gravitational redshifts between
the centre and edges are in the order of $\sim10\;\mathrm{km/s}$. We 
further find in our work here that the signal is also sensitive to 
the nature and depth of the subhalo potentials associated with 
galaxies. We
experiment with different potentials for the subhalos 
and ways to compute the mean gravitational redshift
contribution associated with a particular galaxy (see
Sec.~\ref{sec:halos} for more details). Our fiducial
computation takes  the mean potential
for all particles within each subhalo and uses this potential to
compute gravitational redshifts of the galaxy in 
the subhalo. In Sec.~\ref{sec:halos} we experiment with other prescriptions.

 The difference of  gravitational redshifts
for a galaxy pair $g1-g2$  is given
by

\begin{equation}\label{eq:grav}
\delta z_\mathrm{g} = z_\mathrm{g2}-z_\mathrm{g1} =-\left(\Phi_2-\Phi_1\right)/c^2,
\end{equation}
where $\Phi_2$ and $\Phi_1$ are the gravitational potentials
with respect to infinity for $g1$ and $g2$.

\subsection{Relativistic Doppler effect}
\label{sub:td}

The relativistic Doppler effect has two components, one
longitudinal and the other  transverse. The longitudinal
Doppler effect is dependent on the line-of-sight relative 
motion between the source
and the observer and is affected by both Lorentz contraction and 
wavefront distortion. In the non-relativistic limit the longitudinal
Doppler effect provides an additional redshift increment $\delta
z=\sqrt{{(1+\beta_\mathrm{z})}/{(1-\beta_\mathrm{z})}}-1\approx \beta_\mathrm{z}$ where $\beta_\mathrm{z}$
is the line-of-sight velocity (with positive values signifying
 motion away from the observer). This
effect is included in the usual mapping of galaxy positions
to redshift-space as being the effect of peculiar velocities.

 The second component is the
transverse Doppler effect, and is not usually included in calculations
(the first application was due to \citealt{zhao2013}).
It is observed when the source is moving across
the line-of-sight. The received frequency is therefore reduced by the
Lorentz factor due to the time dilation predicted by  Special
Relativity. As a result, there is an additional redshift increment
$\delta z = \gamma - 1\approx |\bm{\beta}|^2/2$. Note that the
transverse Doppler effect is a second order effect in terms of
$\beta$. In galaxy clusters, $|\bm{v}|\approx 800\;\mathrm{km/s}$,
then $c|\bm{\beta}|^2/2\approx1\;\mathrm{km/s}$, which is less than
but still comparable to the gravitational redshift effect. In our
simulation, a pairwise difference of $g1-g2$ in TD effect is given by
\begin{equation}\label{eq:td}
\delta z_\mathrm{TD} = z_\mathrm{TD2}-z_\mathrm{TD1} =\left(|\bm{\beta}|^2_2-|\bm{\beta}|^2_1\right)/{2}.
\end{equation}

As has been studied by \cite{zhao2013} and K13, the
redshift asymmetry caused by the
transverse Doppler effect has the opposite sign to the gravitational
redshift. If galaxies are all in virialised regions, the TD effect
would be half the  magnitude of the gravitational redshift term
because of the Virial Theorem:
\begin{equation}
\label{eq:virial}
\langle z_\mathrm{TD}\rangle = {\langle|\bm{\beta}|^2\rangle}/{2} = \left\langle{GM}/{r} \right\rangle/{2c^2} = -{\langle\Phi\rangle}/{2c^2} = {\langle z_\mathrm{g}\rangle}/2.
\end{equation}

In our case, looking at the entire galaxy population we expect
the TD effect to be somewhat less important.

\subsection{Light cone effects}

An additional asymmetry in redshift-space clustering is caused 
by light cone (LC) effects.
Following K13, light cone asymmetry originates from the
distortion of phase space density due to our observations of
galaxies on our
past line cone. This leads to the observation
of more galaxies moving towards  us
than moving away from us.  The evolution of the Universe from the high
redshift regime to the low redshift regime, also leads
to an asymmetry. In phase space,
there exists a non-trivial Newtonian transformation of phase space
density from rest-frame (RF) to LC coordinates
$\rho_{\mathrm{LC}}(\bm{r},\bm{\beta})=(1-\beta_\mathrm{z})\rho_{\mathrm{RF}}(\bm{r},\bm{\beta})$. In
K13, the resulting increment of the redshift 
of galaxies in galaxy
clusters was shown to be
\begin{equation}
\begin{split}
\langle z_\mathrm{LC}\rangle& = \int\mathrm{d}x\int\mathrm{d}^3\beta\left(\rho_\mathrm{LC}-\rho_{\mathrm{RF}}\right)(\bm{r},\bm{\beta})(-\beta_\mathrm{z}+\beta^2/2-\Phi/c^2)\\
& = \int\mathrm{d}x\int\mathrm{d}^3\beta\rho_{\mathrm{RF}}(\bm{r},\bm{\beta})(\beta_\mathrm{z}^2-\beta_\mathrm{z}\beta^2/2+\beta_\mathrm{z}\Phi/c^2).\end{split}
\end{equation}

For each individual galaxy, the LC effect boosts the redshift by
\begin{equation}
\label{eq:lc1}
z_\mathrm{LC}=\beta_\mathrm{z}^2-\beta_\mathrm{z}\beta^2/2+\beta_\mathrm{z}\Phi/c^2=\beta_\mathrm{z}^2+O(\beta_\mathrm{z}^3)\approx \beta_\mathrm{z}^2,
\end{equation}
here we only keep the first term as $\beta^2$ and $\Phi/c^2$ 
enter at the same order in $\beta$ as $\beta_\mathrm{z}^2$, following our
discussion in Sec.~\ref{sub:grav} and Sec.~\ref{sub:td}. Using the above
approximation, the pairwise difference in LC coordinates is
\begin{equation}\label{eq:lc}
\delta z_\mathrm{LC} = z_\mathrm{LC2}-z_\mathrm{LC1} ={\beta}^2_{2,\mathrm{z}}-{\beta}^2_{1,\mathrm{z}}.
\end{equation}

If we compare Eqn. \ref{eq:lc} with Eqn. \ref{eq:td}, we
can see that the LC effect is
smaller than TD effect but stays at the same order of magnitude.
We note that $z_\mathrm{LC}=2z_\mathrm{TD}/3$ for isotropic galaxy motions.

\subsection{Special relativistic beaming effect}

Galaxy luminosities and spectra are affected by relativistic beaming, due
to their peculiar motions. The direction and magnitude of galaxy
peculiar motions can therefore influence whether galaxies lie inside
or outside magnitude and colour cuts used to select 
galaxy redshift survey targets from photometric samples
(K13). In the simplest
case, galaxies moving towards the observer (for example if they are on 
the far side of a mass concentration) will be brighter than galaxies
moving away from the observer (on the nearside of a mass concentration).
This will lead to relatively more of one type of galaxy observed
than the other, and therefore an asymmetry in the observed structure.

In our analysis we  average the redshift pairwise
differences  with equal weight per observed galaxy,
this does not have to be always the case. For example galaxies can be weighted
according to their 
luminosities, which would directly be influenced by relativistic beaming.

In practice, the magnitude and even the sign of relativistic beaming 
depends on details of the galaxy spectra as well as their
peculiar velocity. When modeling the effect, for simplicity, we introduce 
a weight for each galaxy in our simulated sample. This weight ($w_{\rm beam}$
below) is given by
%\begin{equation}
%\label{eq:srb}
%\mathrm{Prob.} =\begin{cases}
%1-{\beta_\mathrm{z}}\times\texttt{beamfac}  & \text{for } \beta_\mathrm{z}>0\\
%1 & \text{for } \beta_\mathrm{z} \leq 0\end{cases}
%\end{equation}
\begin{equation}\label{eq:srb}
w_{\rm beam} = 1-\beta_\mathrm{z}\times f_{\rm beam},
\end{equation}
where $f_{\rm beam}$ is a factor which is computed from
the spectral characteristics of the galaxy population.

Treating $w_{\rm beam}$ as the probability of
including a galaxy in the sample, we can see that Eqn.~\ref{eq:srb} indicates 
that we observe all galaxies moving towards us
and a fraction of galaxies moving away from us, depending on their
line-of-sight velocities. This model is linear in $\beta_\mathrm{z}$, and
we have found that it behaves robustly when computing the 
redshift asymmetry using the defined ``shell estimator''. We use 
$w_{\rm beam}$ to weight galaxies before computing their clustering.

The value of $f_{\rm beam}$ depends on the characteristics of the
galaxy observational sample we are modeling.
A given galaxy catalogue is affected by  relativistic beaming when 
the sample is selected with cuts in colour-magnitude space,
 because the observed flux of galaxies is altered due to 
relativistic beaming. K13 made some general
 assumptions about the spectra 
of galaxies used in the \cite{wojtak2011} cluster measurement
and assumed the simplest case of a flux-limited sample to derive 
$f_{\rm beam}=6$. 
The photometric selection of galaxies in many redshift samples
is more complex, however, and to predict the effect  of  beaming with
 more accuracy it is necessary to model
the specific population of  galaxies and selection criteria in detail.

In a companion paper (\citealt{Alam2016TS}) we therefore
explore the impact of relativistic beaming on the selection
of galaxies in the SDSS/BOSS/CMASS redshift sample. We use the observed 
spectra as  templates to model SRB and define $w_{\rm beam}$ as the ratio 
of the number of galaxies passing the target selection criteria 
before and after SRB is applied to their spectra. 
We find that the
spectra and hence values of $f_{\rm beam}$ for each galaxy
depend on galaxy stellar mass and redshift. 
We therefore 
compute $w_{\rm beam}$ as a function of $\bm{\beta}$, stellar mass 
and redshift \citep[see Fig.~5 in ][]{Alam2016TS}. 
In order to apply the results of this analysis to the simulations
in this paper, we compute a representative value of $w_{\rm beam}$ as 
a function of  line-of-sight velocity $\beta_\mathrm{z}$ only, by 
 weighting the individual $w_{\rm beam}$ values by the number of galaxies 
in the corresponding redshift and stellar mass bin in the SDSS/BOSS/CMASS
sample. 

\begin{figure}
  \centering
    \includegraphics[width=0.5\textwidth]{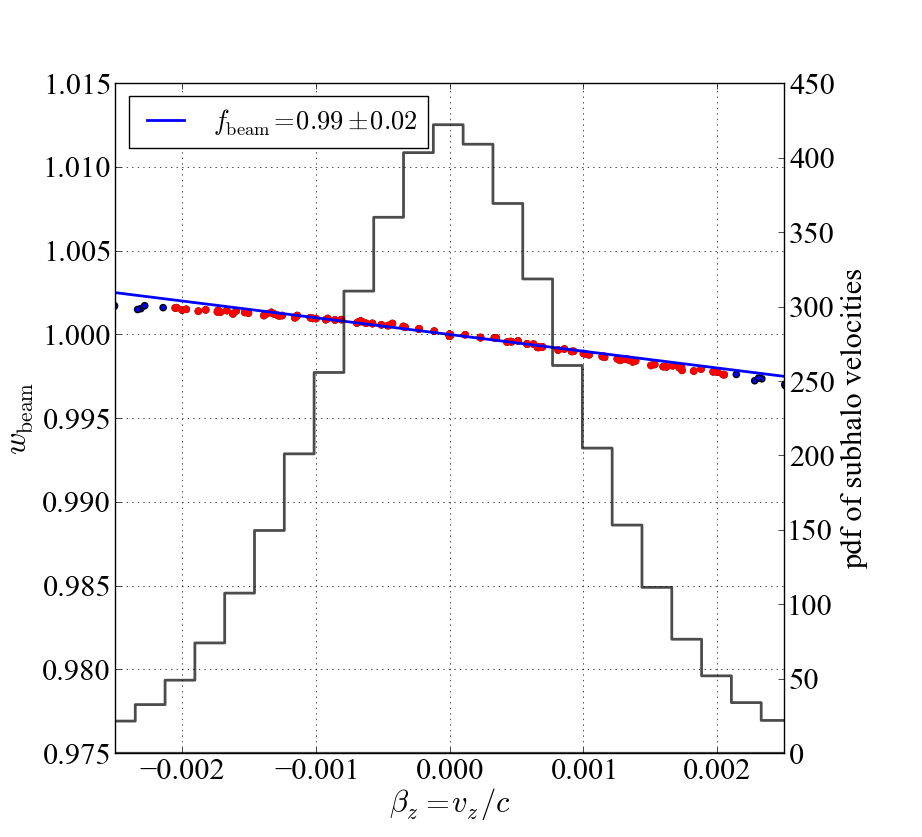}
     \caption{ The effect of  relativistic  beaming on the probability
of including a galaxy in the SDSS/BOSS/CMASS galaxy redshift sample as 
a function
of line-of-sight peculiar velocity. The dots show the probability of
including galaxies computed after marginalizing over galaxy stellar
mass and redshift using the distribution of data in the observed 
sample (\citealt{Alam2016TS}).  The red dots show data within a 
velocity range of  -630 km/s and 630 km/s, which encompasses 95\%
of the galaxies in our simulations. 
The blue line is a fit to the red dots using the
functional form of Eqn.~\ref{eq:srb}.
The distribution
of line of sight galaxy velocities in one of our simulation realizations
is shown as a black histogram. }
     \label{fig:beaming_fitting}
\end{figure}

 We find for
the purpose of estimating the impact of SRB on clustering asymmetry
that  Eqn.~\ref{eq:srb}, which is only a linear
function of line-of-sight velocity
is general enough to adequately model the effect.
In Fig.~\ref{fig:beaming_fitting} we show the
the pdf of subhalo velocities from one of our
simulation realizations to illustrate the limited range of
peculiar velocities that are relevant. We also plot the $w_{\rm beam}$
values  from SDSS/BOSS/CMASS redshift sample 
(see \citealt{Alam2016TS}), and the linear fit of Eqn.~\ref{eq:srb}
 over the range -630 km/s and 630 km/s. The best fitting value of
$f_\mathrm{beam}$  is 1.0 and we use this value 
in the rest of this paper.

\subsection{Luminosity distance perturbation} As was
    recently reviewed in \cite{kaiser2015}, one of the first-order
    corrections in terms of peculiar velocities is the luminosity
    distance perturbation. The inferred luminosity of each galaxy,
    which is inversely proportional to the square of the luminosity
    distance, is perturbed by the effect of peculiar velocities on the
    redshift used to calculate that luminosity distance. For
    simplicity, in our simulations, we assume mass traces light.
    Under such an assumption, since the flux-distance relation follows
    an inverse-square law, we can model the mass perturbation in
    observations using
    Eqn.~\ref{eq:ldp}.  \begin{equation}\label{eq:ldp}
      M_\mathrm{obs}=M\left(\frac {d(z)}
        {d(z)+\beta_\text{z}c/H(z)}\right)^2, \end{equation} where
    $d(z)$ is the comoving distance at redshift $z$ (proportional to
    the luminosity distance) and $\beta_\text{z}c/H(z)$ is the
    perturbation caused by the peculiar velocity. This pertubation is
    usually small for SDSS/BOSS/CMASS galaxies at $z=0.57$ (of order a
    part in one thousand).  It can be seen that the lowest order
    correction is first-order in the peculiar velocity.

    A
    discussion of the first-order velocity contributions can also be
    found in Eqn. 28 of \cite{mcDonald2009}, Eqn. 3 of \cite{yoo2012}
    and Eqn. 29 of \cite{bonvin2014}. For example, Eqn. 3 of
    \cite{yoo2012} gives the luminosity distance contribution as
    $-5p\delta D_L$ where $p$ is the slope of the galaxy
luminosity function. However in the  discussion in Sec. III of
\cite{yoo2012}  it is claimed that
    first-order relativistic contributions are cancelled, and therefore the
    Newtonian calculation can fully reproduce Eqn. 3 of that paper. In our
    paper, we show how  first-order and second-order
    relativistic terms contribute to the
    signal. The leading first-order relativistic terms are 
    those due to special
    relativistic beaming and the luminosity distance perturbation.  Other
    first-order relativistic terms are also discussed in
    Sec.~\ref{sec:others}. The combination of these terms 
    covers the first-order relativistic effects, but in our case as we 
consider non-linear scales, second-order effects are also important.

\subsection{Wide-angle effects}

In this paper we have so far assumed the distant observer 
approximation however this
will lead to errors the case of the SDSS/BOSS/CMASS survey. The non-parallel
lines-of-sight to galaxies will also induce asymmetries
in the 2d cross-correlation functions leading to corrections to the
multipoles and shell estimators. From Fig.~12 of \cite{Gaztanaga2015},
it can be seen that 
the wide-angle effects are of the same order of magnitude as the
relativistic effects on the $r > 20 $ Mpc/$h$  scales dealt
with in that paper. We compute the correction due to   wide-angle
effects following \cite{Gaztanaga2015} (see our companion paper
\citealt{Giusarma2016PT}) for a more detailed treatment and add them to 
the shell estimator we
derived from the distant observer approximation. In Fig.~\ref{fig:est_shadab} we
plot $z^{\rm shell}$, with and without the wide-angle
correction. It can be seen that on the small scales
which are most important for our current work the wide-angle effect is 
insignificant.

\subsection{Other effects}
\label{sec:others}
In the context of galaxy clusters, \cite{cai2016} recently 
derived the total redshift of a galaxy relative to a 
stationary source in the observer's past light cone.
They found that alongside terms that are functions of the peculiar velocity,
some additional cross terms eppear (see their Eqn. 12).
These cross terms are $-zg_\mathrm{z}$ and $Hzv_\mathrm{z}/c$, 
where $-zg_\mathrm{z}$ is minus the 
product of the line-of-sight displacement with the gradient of 
potential in the line-of-sight direction, and $Hzv_\mathrm{z}/c$ 
is the product of the line-of-sight Hubble velocity with the line-of-sight 
peculiar velocity. In their simulated cluster-galaxy clustering,
  on scales $<10$ Mpc/$h$, \cite{cai2016} find
these terms to be as small as the TD effect,  while on 
large scales both terms should vanish because the individual
 quantities that enter into the cross terms are 
no longer correlated. We also investigate these two terms in our 
simulation as follows.

For these two terms, we compute the boosts for each individual subhalo,
\begin{equation}
z_\mathrm{cross}=-zg_\mathrm{z}+Hzv_\mathrm{z}/c.
\end{equation}
The pairwise difference is then
\begin{equation}
\begin{split}
\delta z_\mathrm{cross}&=z_\mathrm{cross2}-z_\mathrm{cross1}\\
&=z_1g_\mathrm{1,z}-z_2g_\mathrm{2,z}+Hz_2\beta_\mathrm{2,z}/c-Hz_1\beta_\mathrm{1,z}/c,
\end{split}
\end{equation}
where $z_1$ and $z_2$ are line-of-sight distance in real space.

\section{Simulation}
\label{sec:simu_tests}
We have run 8 cosmological $N$-body simulation realizations using the
$\textsc{P-Gadget3}$ code. The cosmological parameters used are summarised
in Table~\ref{tab:cosmo_params}. In our fiducial
 runs,  we have used  $1024^3$ particles in a
cubic periodic box with side length $1\;\mathrm{Gpc}/h$. The mass per
particle is $7.13\times10^{10}\;M_\odot/h$, and the gravitational force
softening length is $20\;\mathrm{kpc}/h$. Each simulation starts at
$z=159$ and runs till $z=0$. $\textsc{SubFind}$
(\citealt{springel2001subfind}) is used to identify galaxy-sized
subhalos. As we mentioned in Sec.~\ref{sub:grav}, we take the 
potential of all stars in a galaxy in a 
 subhalo to be equal to the mean of the potentials of
all the particles in the subhalo. 
%The potentials are relative to the
%mean level in the universe.

\begin{table}
 \centering
  \caption{Cosmological parameters used in $N$-body simulations}
  \begin{tabular}{rrrrrr}
  \hline
  $h$ & $\Omega_\mathrm{baryon}$ & $\Omega_\mathrm{cdm}$ & $\Omega_\Lambda$ & Spectral index & $\sigma_8$\\\hline
 0.7 & 0.0462 & 0.2538 & 0.7 & 0.96 & 0.873285\\\hline
\end{tabular}
\label{tab:cosmo_params}
\end{table}

\subsection{Galaxies and large-scale structure}

\begin{figure*}
  \centering
    \includegraphics[width=0.7\textwidth]{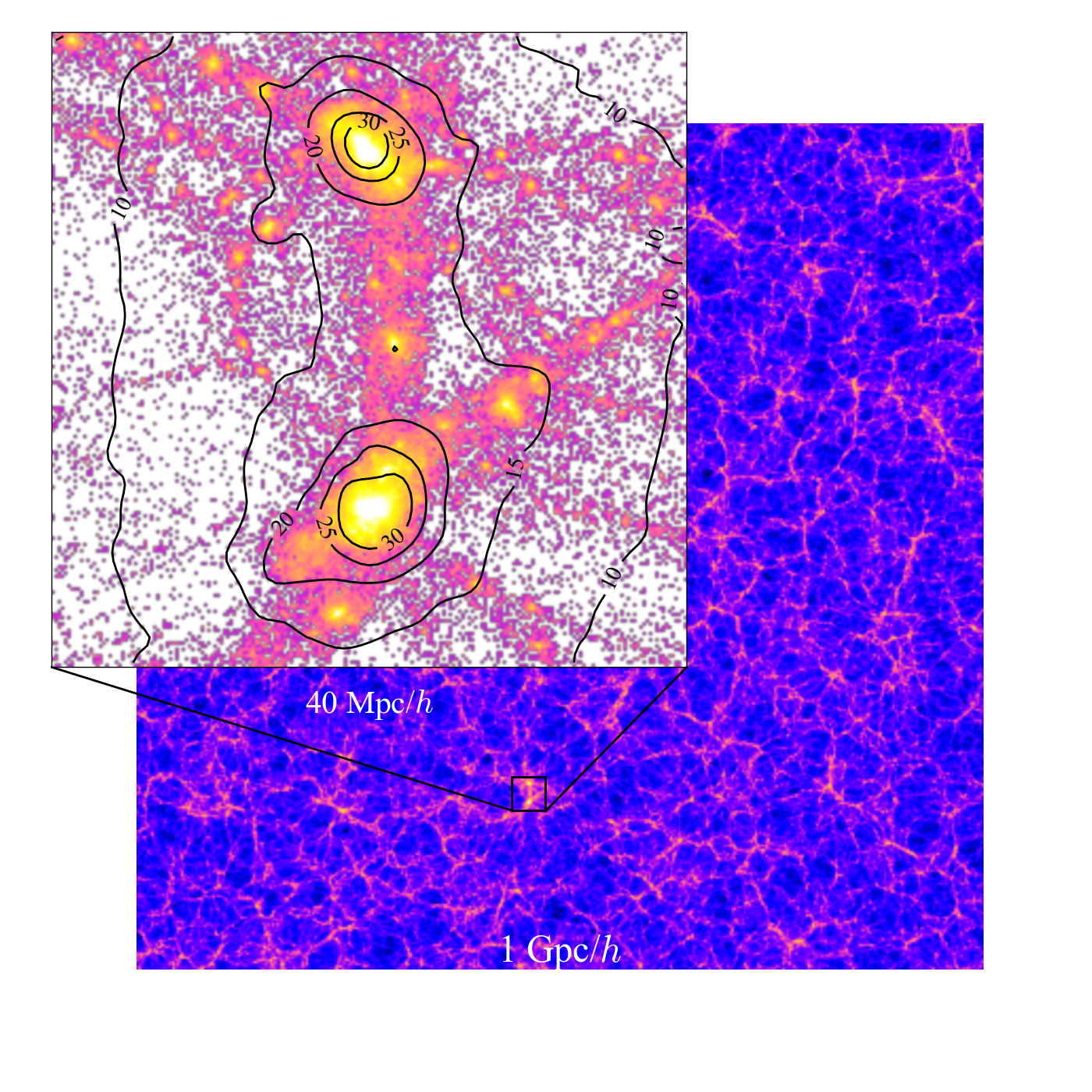}
     \caption{Dark matter distribution with gravitational redshift contours. We show a $10\;\mathrm{Mpc}/h$ thick slice of a cubic box with side length $1\;\mathrm{Gpc}/h$ at $z=0.57$. We also magnify a matter-abundant region $40\times40\times10\;\left(\mathrm{Mpc}/h\right)^3$ and show its density as well as gravitational redshift contours on the top left panel. The numbers on the contours are in unit of km/s.}
     \label{fig:sliceplot}
\end{figure*}

In Fig.~\ref{fig:sliceplot} we show
 an illustrative slice  through  one of the simulation realizations
at $z=0.57$.  The gravitational potential (redshift) contours are
much smoother than the  density distribution. The gravitational
redshift differences between the centres and edges of the massive 
(cluster-sized) halos shown in the inset are of order
$\sim10\;\mathrm{km/s}$ as expected. Because the simulations have no model for 
galaxy formation, we associate galaxies
with dark matter subhalos. The more massive galaxies tend to
reside inside deeper potential wells while the less massive ones are
in shallower potential wells on average. This leads to the difference in
$z_\mathrm{g}$ when averaged over $g1-g2$ galaxy pairs.

\subsection{2d cross-correlation function}

We  compute the cross-correlation function between two
subsamples of galaxies (as in C13 and \citealt{bonvin2014}).
To create the subsamples, we  
apply a mass cut to the subhalos in each realization, splitting 
the subhalo catalogue into two equal number halves.
When analyzing observational data, it is usually simplest to 
apply a luminosity cut, but in our simulations we use
mass to divide the samples. In Sec.~\ref{sec:bias} we explain how we deal
with the mapping between mass and luminosity in order to compare
to observations.

 To compute the cross-correlation function in redshift space,
which includes the relativistic distortions described in 
Sec.~\ref{sec:modeling} we first rescale the mass of each galaxy according to LDP, take each galaxy pair, add three
 components (the gravitational redshift effect, the TD effect and the LC effect),
due to the difference in real space $z$, the difference in peculiar
velocity and the sum of terms computed from Eqn.~\ref{eq:grav}, ~\ref{eq:td},
~\ref{eq:lc} and add the wide-angle contribution computed from our companion paper \cite{Giusarma2016PT}.  We do not model the relativistic beaming effect
directly from redshift differences between galaxy pairs. Instead, we weight
each galaxy by the probability given in Eqn.~\ref{eq:srb}.
As the relativistic effects are extremely small, we scale up 
each by a large factor (by multiplying the subhalo velocity that goes 
into computing each effect by this factor).
This enables us to compute the asymmetries
in clustering from the simulations with much greater signal to noise,
where the statistics are of course appropriately renormalised by the
scaling factor after being measured.
The choice of factors and tests for convergence are
 discussed in more detail in Sec.~\ref{sec:shell_dipole}.

\begin{figure}
  \centering
    \includegraphics[width=0.5\textwidth]{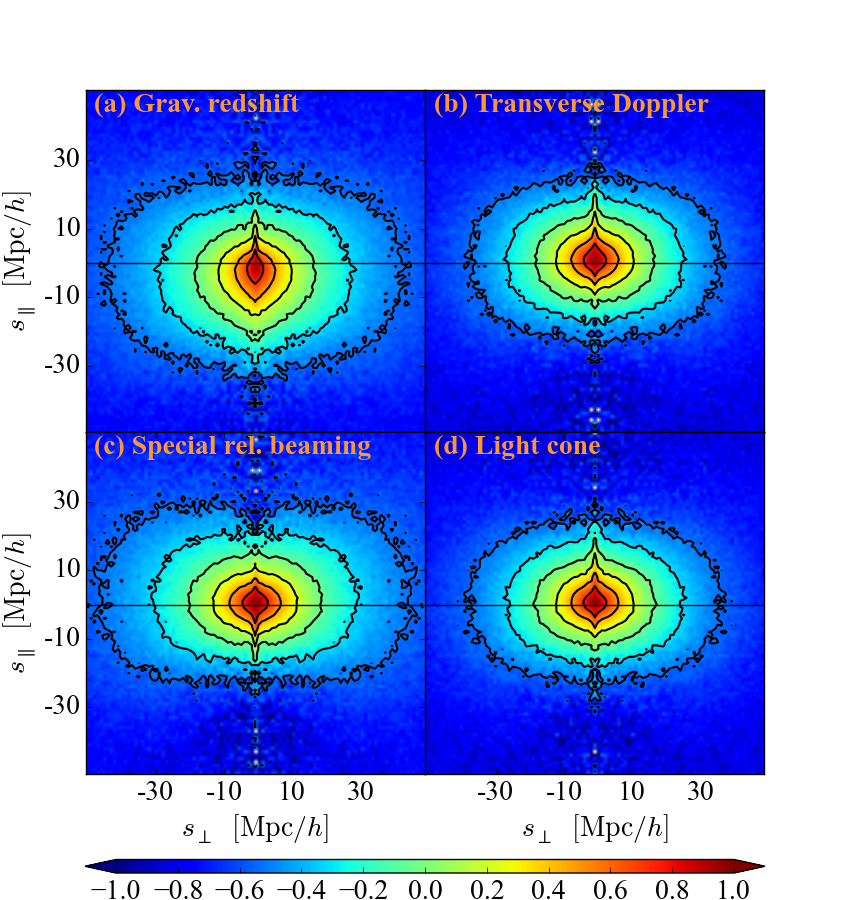}
    \caption{The 2d 2-pt cross-correlation of the $g1-g2$ galaxy
      samples for different relativistic effects in the redshift space (plots made using one simulation realization). We exaggerate all the effect by multiplying them with a constant factor of 250. Top left: With Grav. redshifts
      $\times250$. Top right: With TD effect $\times250$. Bottom left:
      With SRB effect $\times250$. Bottom right: With LC effect
      $\times250$. All four effects are computed at $z=0.57$ with the mass
threshold for the $g1$ (high mass) sample being
      $M_\mathrm{subhalo} > 10^{13}\;M_\odot/h$. In each panel, the
      colourmap and contours show the distortion of 2-pt
      cross-correlation function. We have chosen not to plot the LDP and the wide-angle effects here, as LDP is quite similar in effect to SRB (see Fig.~\ref{fig:shell_contri}) and the wide-angle effects are computed 
independently of the simulations.}
     \label{fig:corr2d}
\end{figure}

In Fig.~\ref{fig:corr2d} we show the cross-correlation function
of the $g1$ and $g2$ galaxy samples, as a function of separation 
parallel and perpendicular to the line of sight. For these 
illustrative plots, we scale up the relativistic effects by a factor
of 250, in order to make them visible. 
We show the cross-correlation function
for each of the four relativistic effects individually. The four effects
cause line-of-sight asymmetries with different forms and magnitudes.
Comparing the panels we can see that the gravitational redshift and the SRB
effect are the dominating two effects. The gravitational redshift
effects tends to drag the cross-correlation function bluewards,
causing a relative blueshift for the low-mass galaxy ($g2$) with respect to
high-mass galaxy ($g1$). This makes sense because low-mass galaxies
are more loosely gravitationally bound so that they have smaller gravitational
redshifts. The SRB effect causes a ``flattening'' of the
cross-correlation function. The asymmetries due to the SRB, TD and LC
effects are difficult to see by eye, but as we show below can 
be quantified
using multipoles and the shell estimator. Note that we do
not plot the LDP and 
the wide-angle effects here. This is because the LDP behaves very similarly to 
SRB and the wide-angle effects are quite dependent on the bias differences 
between the two galaxy samples (see Sec.~\ref{sec:bias}).

\subsection{Dipole}\label{sec:dipole}

We decompose the 2d cross-correlation functions into different moments  
using Legendre polynomials. 
\begin{equation}
\xi_l(r)=\frac {2l+1} 2\int\limits_{-1}^1\,\xi(r,\cos\theta)P_l(\cos\theta)\,\mathrm{d}\cos\theta,
\end{equation}
where $l=0,1,2$ correspond to the monopole, dipole and quadrupole
respectively, and $P_0(\cos\theta)=1, P_1(\cos\theta)=\cos\theta, 
P_2(\cos\theta)=(3\cos^2\theta-1)/2$ are the first three Legendre polynomials. The monopole corresponds 
to the isotropic part of the cross-correlation functions. 
%Both the dipole and 
%the quadrupole
%quantify asymmetries but in different ways.
The quadrupole quantifies the anisotropy in the correlation 
function in the directions parallel to the  line-of-sight and perpendicular to the line-of-sight. The dipole quantifies the asymmetry in 
the correlation function with respect to  positive and negative 
line-of-sight separations. 
As we can see from Fig.~\ref{fig:corr2d}, by 
carrying out a Legendre decomposition, we could quantify the 
line-of-sight asymmetries for all four effects using the dipole.
The gravitational redshift leads to a relative blueshift
 for low mass galaxies compared to high mass galaxies, which causes 
a negative dipole on small scales.
On the other hand, on large scales, the correlation function is 
redshifted along the line-of-sight and more flattened, and 
we shall see that this leads to a measurable effect of opposite sign.
The SRB and LDP effects blueshift the correlation function along 
the line-of-sight at scales of a few
$\mathrm{Mpc}/h$ or less, but redshifts and compresses the correlation
function on larger scales. The asymmetries from the TD and LC effects
are both smaller, but both lead to a positive dipole (opposite sign
to the  gravitational redshift) on small
scales. To quantify the asymmetries we find that 
the dipole is not practical in noisy simulations (see Sec.~
\ref{sec:shell_dipole} for discussion of this point), but instead
we use a shell estimator.

%We show the dipoles (red) and
%the quadrupoles (black) in Fig.~\ref{fig:legendre}, again
%with the different effects added separately in different panels. By
%comparing them, we find the
%gravitational redshift leads to a relative blueshift
%(negative dipole) and causes the
%correlation function to be more stretched along the line-of-sight on small
%scales. On the other hand, on large scales, the 
%correlation function is redshifted
%along the line-of-sight and more flattened. The SRB effect blueshifts
%the correlation function along the line-of-sight at a few
%$\mathrm{Mpc}/h$, heavily redshifts and squashes the correlation
%function at large scales. The asymmetries from the TD and LC effects
%are smaller. Both of them lead to a positive dipole (opposite sign
%as gravitational redshift) on small
%scales.
%
%\begin{figure}
%  \centering
%    \includegraphics[width=0.5\textwidth]{./plots/correlation_legendre.png}
%    \caption{Projection of 2d cross-correlation functions in the
%      Legendre basis $l=0,1,2$. The dashed lines are the multipoles in
%      the redshift space without any relativistic effects (only peculiar 
%velocities)
% and the solid lines are
%      the corresponding effect $\times250$. The blue, red and black
%      lines show monopoles, dipoles and quadrupoles respectively. The
%      error bars are errors on the mean of 8 realizations.}
%     \label{fig:legendre}
%\end{figure}

\subsection{Shell estimator}\label{sec:shell}

Another means to make relativistic distortions of clustering
measurable is to use the shell estimator
which is designed to probe the line-of-sight asymmetry in the cross 
correlation. C13 designed the estimator to be close in physical meaning
to the gravitational redshift profiles measured in 
galaxy clusters (e.g., by \citealt{wojtak2011}).
In the shell estimator we bin pairs of subhalos according to their
 pair separation and average the line-of-sight component of the separation, weighting the pairs by weights from SRB (Eqn.~\ref{eq:srb}) in which the weights fluctuate around 1, effectively computing the redshift displacement of the centroid as a function
of pair separation. The mathematical form is Eqn.~\ref{eq:shellsim}:
\begin{equation}\label{eq:shellsim}
z^\text{shell}(s) = \frac {H\sum\limits_{\alpha \in g1, \beta \in g2}^{|r_{\alpha\beta}-s|<\Delta s}(z_\alpha - z_\beta)w_{\rm beam, \alpha}w_{\rm beam, \beta}} {\sum\limits_{\alpha \in g1, \beta \in g2}^{|r_{\alpha\beta}-s|<\Delta s}w_{\rm beam, \alpha}w_{\rm beam, \beta}},
\end{equation}
which is also mentioned in our companion paper (\citealt{Alam2016Measurement}).
 This estimator was 
also used by \cite{Gaztanaga2015} and \cite{Alam2016Measurement}. In the measurement with SDSS/BOSS/CMASS samples, the shell estimator is computed from the correlation function and it takes the form of
\begin{equation}
z^\text{shell}(s)=\frac {H\int\limits_{0}^{\pi}\text{d}\theta's'_\parallel\left[1+\xi\left(s',\theta'\right)\right]} {\int\limits_{0}^{\pi}\text{d}\theta'\left[1+\xi\left(s',\theta'\right)\right]},
\label{zshell}
\end{equation}
where the integral over $\theta'$ averages the line-of-sight pair difference $s'_\parallel$ in each radial bin.

\begin{figure}
  \centering
    \includegraphics[width=0.5\textwidth]{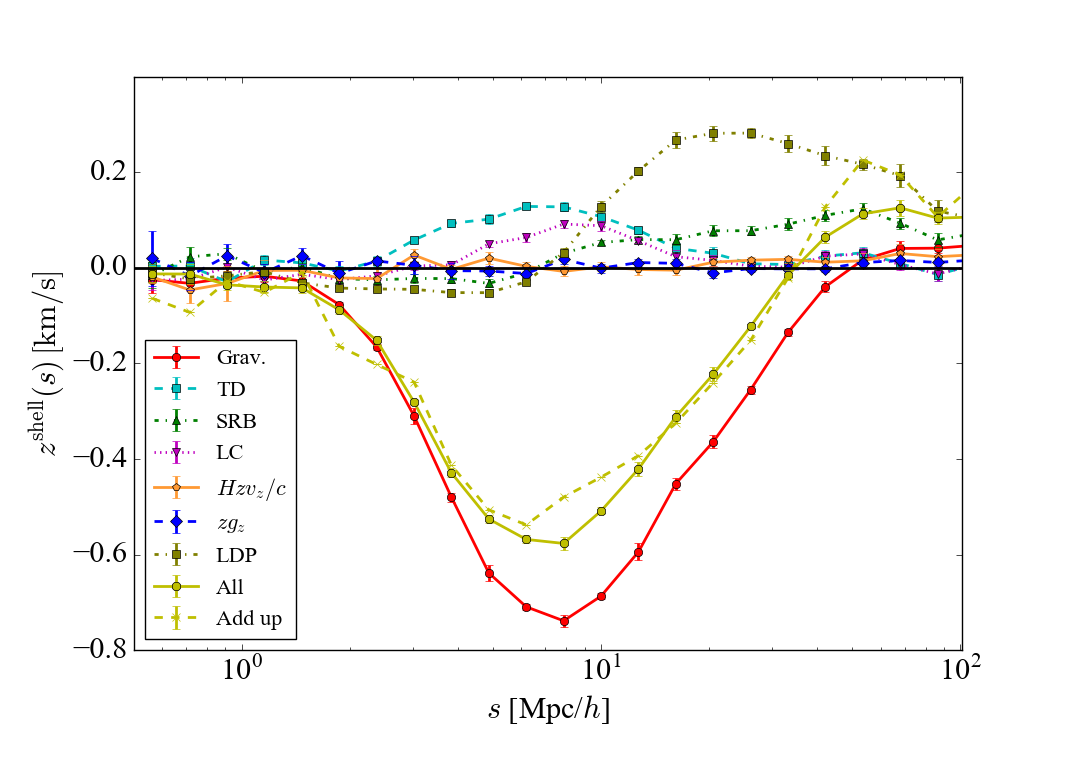}
    \caption{The shell estimator Eqn.~\ref{zshell}, of clustering asymmetries
computed from
the cross-correlation function
of $g1-g2$ galaxy samples. We show
 the different relativistic effects individually:
the gravitational redshift, TD, SRB, LC, $Hzv_z/c$, $zg_z$ and LDP. 
The curves are the average of results from 8 realizations at $z=0.57$ with 
      masscut $3\times10^{13}\,M_\odot/h$. The error bars show the
      errors on the mean. The solid yellow line shows
 $z^\text{shell}$  for the five significant effects (the gravitational redshift, TD, SRB, LC, LDP) included at once
 and the dashed yellow line is the
      sum of the curves for these five effects. Here a boost factor 100 is applied. How  the boost factor is chosen is discussed in Sec.~\ref{sec:shell_dipole}}.
     \label{fig:shell_contri}
\end{figure}

Note that $z^\text{shell}$ is zero in real space and
 $z^\text{shell}(s) \rightarrow 0$ as $s \rightarrow 0$
even in redshift space. Fig.~\ref{fig:shell_contri} shows the
contributions to the shell estimator from the four different
relativistic effects. Among them, the gravitational redshift effect is the most
significant one, causing a relative blueshift for galaxy pairs 
on small scales, which peaks when 
 $s \sim 8$ Mpc/$h$. The LDP term is the largest in magnitude on scales $s > 20$ Mpc/$h$. The SRB effect behaves similarly to the LDP term but with a smaller magnitude. Also, the LDP and SRB effects do change sign, at 
$s \sim 8$ Mpc/$h$ -- there is a relative blueshift at small scales and a relative
redshift at large scales.  Although there is a quite complex
 interaction between galaxy spectra and SRB 
(see \citealt{Alam2016TS}), one can approximately explain this pattern
as being due to infall on large scales and  virialised
motions on small scales. The mean relative line-of-sight
peculiar velocity of  pairs of galaxies with given redshift separation
$s$ will therefore change sign as we move between these two regimes, and this
leads to a change of sign in $z^\text{shell}$.

The next  effects in order of significance seen in Fig.~\ref{fig:shell_contri}
are the TD and the LC effects. They result in a relative
redshift at all scales since high mass galaxies move more slowly
on average than
the lower mass member of a pair. The least significant effects
 are the cross terms $Hzv_z/c$, $zg_z$ discussed in \cite{cai2016}. 
Our $g1-g2$ galaxy cross-correlation almost shows no sign of these two
effects. This is expected as both subsets $g1$ and $g2$
contain equal numbers of galaxies, which leads 
to  small correlations in $z$, $v_z$ and $g_z$ when averaging 
over populations of relatively similar objects. These effects are even
smaller that for the cluster-galaxy cross-correlation case 
simulated by \cite{cai2016}, where they were also not significant 
compared to the gravitational redshift.

We also show in  Fig.~\ref{fig:shell_contri} the
combined effect (``all'') which we obtain by adding each of five 
effects (the gravitational redshift, TD, SRB, LC and LDP) to the 
subhalo positions in redshift space. The curve labelled
``add up'' is the sum of these individual five effect curves (computed by 
measuring $z^\text{shell}$ from galaxy catalogues with only one
relativistic effect included at a time). The two curves
 (``all'' and ``add up'') agree quite well,
showing that the effects can be added linearly, even on these 
relatively small scales. However, we find taking account of two cross terms 
only adds noise to the overall signal. 
Therefore we only consider the five effects mentioned
while excluding those two cross terms. The 
amplitude of $z^\text{shell}$ while including all the effects 
is dominated by  the gravitational redshift effect
on small scales and the LDP effect on large scales ($s > 30$ Mpc/$h$).

\subsection{Increasing the signal to noise ratio of the
simulation predictions}\label{sec:shell_dipole}

As we mentioned in Sec.~\ref{sec:dipole} and Sec.~\ref{sec:shell},
both the dipole and the shell estimator are capable of quantifying the
signal due to relativistic effects (the antisymmetric part of
the cross-correlation function could also be used- see e.g.,
\citealt{JCAP2016}). We choose the shell
estimator for three reasons. First, the shell estimator shows the 
signal 
in km/s units, which serves as a direct way to express line-of-sight
redshift (velocity) distortions. Second, we have seen in 
Sec.~\ref{sec:modeling} that the asymmetry signal on small scales is
dominated by the gravitational redshift, and this has already been
quantified using the very similar gravitational redshift profile in 
several published works on galaxy clusters.

A third and more practical reason to prefer
shell estimator has to do with the limitations of our
simulations. Since our simulations are only 1 (Gpc/$h$)$^3$ in volume
the signal we measure will be quite noisy due to the intrinsically
asymmetric shapes of large-scale structures not being fully averaged out
in the cross-correlation function.
In order to reduce the relative effect of 
noise we boost each of the relativistic effects by 
multiplying the input peculiar velocity of each halo by 
multiplication factor, and then account for the
multiplication factor in the final measurement. Ideally, using
different multiplication factors should only change the noise
characteristics of the signal without affecting its amplitude and shape. 
In practice, this is something which must be explored with a 
convergence test, as described below.

\begin{figure}
  \centering
    \includegraphics[width=0.5\textwidth]{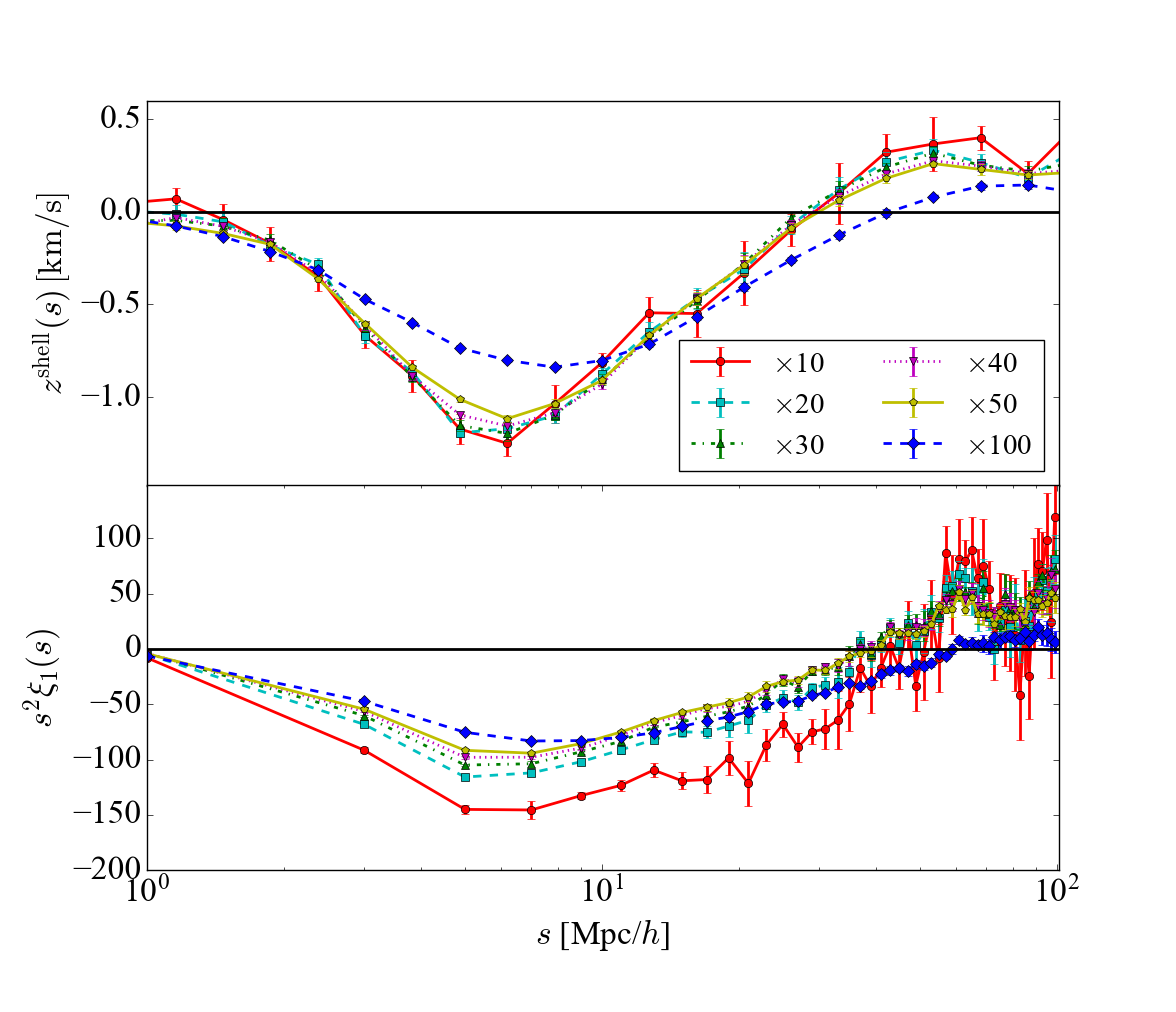}
    \caption{The shell estimator and the dipole moment for different
      multiplication factors. The $x$-axis represents the pair
      separation in the log space. The colour and style schemes are the
      same for both panels. The circles with the solid line (red), the
      squares with the dashed line (cyan), the triangles with the dash-
      dotted line (green), the upside down triangles with the dotted
      line (magenta), the pentagons with the solid line (yellow) and
      the diamonds with the dashed line show the two estimators
      computed using multiplication factors of
      10, 20, 30, 40, 50, and 100 respectively.
       In each case, the peculiar velocities entering in to the calculation
of relativistic effects were multiplied by this factor, and then the
measure statistic ($z^{\rm shell}$ or dipole) was divided by the same factor before plotting (see text). The error bars are the errors on the mean
      of 8 simulation realizations. Note that the shell estimator has a much weaker dependence since the normalization saturates such dependence.}
     \label{fig:shell_dipole}
\end{figure}

We
have performed the measurement of the shell estimator and dipole moment
for a series of different multiplication factors, and the results are
shown in
Fig.~\ref{fig:shell_dipole}. The top panel shows the shell estimator
and bottom panel shows the dipole moment.
 The error bars are obtained from the errors on the mean 
of 8 simulation realization. We observe from the top panel
 that the prediction for the
shell estimator is largely independent of the factor used but the
dipole
moment is highly sensitive to the factor. 
This behavior is likely due to the fact that the
definition of the shell estimator includes a 
normalization in the denominator which 
balances the effect of the boost factor,
something which is not the case for the computation of the dipole moment. To be specific, both the numerator and denominator 
in either Eqn.~\ref{eq:shellsim} or Eqn~\ref{zshell} will fluctuate 
somewhat coherently and mostly cancel or reduce the non-linear
effect when boosting while the dipole moment will not.
We do find that 
shell estimator has some weak dependence on the multiplication
factor,  and that this becomes most noticeable when the boost factor is
increased to 100 or more. When the boost factor is this large
the relativistic effects change the line-of-sight separation 
by significantly more than a bin size. 
This leads to a shifting of the minimum to right and
a  dilution of the
amplitude of the signal. 
%We therefore decide to use the minimum boost factor possible while making sure that we are in a regime with good signal to noise. 
We therefore decide to use the boost factor with good signal to noise while making sure that we are in a regime where the shell estimator is independent of the boost factor. Examining the $z^{\rm shell}$ signal in 
Fig.~\ref{fig:shell_dipole} close to the minimum, we can see that 
a boost factor of 50 leads to a statistical error bar of $\sim 5\%$,
and the systematic error from the behaviour of the convergence at low 
factors is similar or less than this. We therefore choose a boost factor of 
50 in this paper to make our predictions for $z^{\rm shell}$.

\section{Simulating an observed sample of galaxies}
\label{sec:bias}
\subsection{Galaxy bias}
We can study the distribution of galaxies from both
observations and simulations since galaxies are used as tracers
of the matter in the Universe.  Galaxies  are ``biased''
tracers in the sense that the  physics of galaxy formation can 
influence the relationship between galaxies and matter. 
The simplest parametrization of this relationship, galaxy bias
measures the difference between the spatial distribution of galaxies and
the underlying dark matter density field. It is defined as
\begin{equation}
\delta_\mathrm{g} = f(\delta_\mathrm{m}),
\end{equation}
here $\delta_\mathrm{g}=\rho_\mathrm{g}/\bar{\rho_\mathrm{g}}-1$
represents the mean overdensity of galaxies and similarly
$\delta_\mathrm{m}$ represents the mean overdensity of matter. The
function $f(\delta_\mathrm{m})$ depends on both scale and galaxy
evolution. On large scales, the galaxy bias is linear and takes the
following form
\begin{equation}
b = \frac {\delta_\mathrm{g}} {\delta_\mathrm{m}} = \sqrt{\frac {\xi_\mathrm{g}} {\xi_\mathrm{m}}},
\label{biaseq}
\end{equation}
where $\xi_\mathrm{g}$ and $\xi_\mathrm{m}$ are the 2-pt cross-correlation functions of galaxy-matter and matter-matter respectively.

Our primary application of the work in this paper is a prediction
for the clustering asymmetry measured from the SDSS/BOSS/CMASS redshift
sample in our companion paper (\citealt{Alam2016Measurement}). As a result, we would like 
our simulated galaxy catalogue to have the same galaxy bias properties
as the observed sample.
In order to achieve this, we vary two
parameters, the lower limit on the subhalo mass threshold
to be allowed into the sample, and a subhalo mass error 
$\sigma$. The latter is a random variable (normally distributed in $\log(M)$)
which we add to the subhalo masses in order to model the scatter between
the observed quantity (e.g., luminosity) and mass. The subhalo mass after 
adding the scatter is
\begin{equation}
\log_{10} M_{\rm obs,i}=\log_{10} M_{i}+\sigma_{i},
\label{scatter}
\end{equation}
where $\sigma_{i}$ is a normally distributed random variable with
zero mean and standard deviation $\sigma$, and $M_{i}$ is the
actual mass of subhalo $M_{i}$.
 This mass error is adjusted by
tuning the spread of the Gaussian $\sigma$.

\begin{figure}
  \centering
    \includegraphics[width=0.5\textwidth]{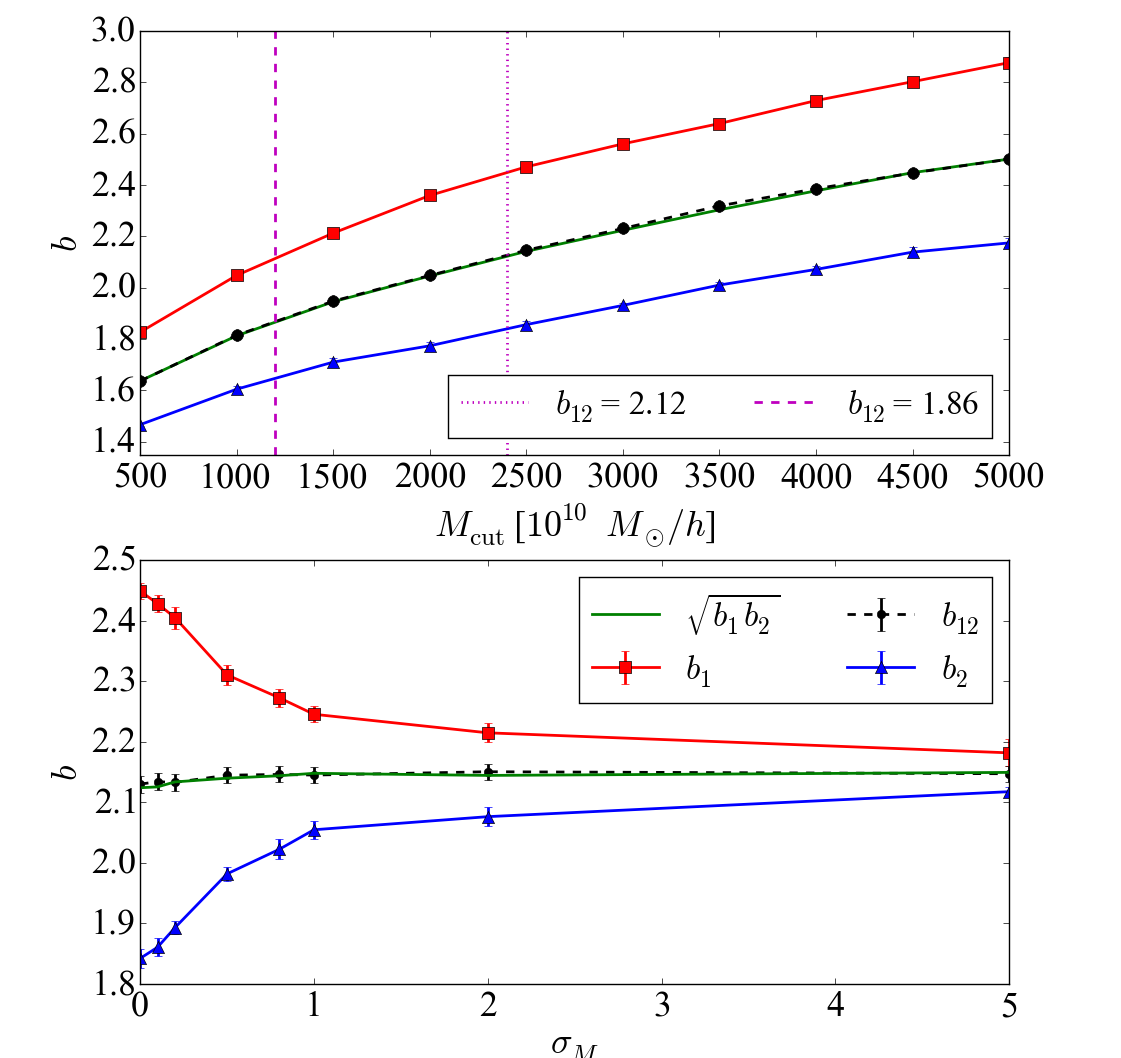}
    \caption{Biases of two galaxy subsets at $z=0.57$ as a function of
      the subhalo mass cut and spread. In both subplots, $b_1$, $b_2$,
      $b_{12}$ are the bias factors of the high mass subhalo subset,
      the low mass subhalo subset and two subsets respectively. The
      solid green curves are the geometric mean of $b_1$ and
      $b_2$. The dashed and dotted magenta line on the top panel indicate the
      mass cut where $b_{12}=1.86$ and $b_{12}=2.12$ respectively.}
     \label{fig:bias_mass}
\end{figure}

As before, we split our subhalo samples into two halves, but this time
using $M_{\rm obs,i}$. We compute the bias of the two samples with respect
to the dark matter by averaging Eqn.~\ref{biaseq} over scales
$r=20-50$ Mpc/$h$. We compute
 results at a simulation output redshift $z=0.57$ which 
is the mean redshift of the CMASS sample (\citealt{Alam2016Measurement}).
 In  the top panel of Fig.~\ref{fig:bias_mass} we
show the effect of a changing mass threshold on the bias $b_{1}$
measured from the high mass half, $b_2$ (low mass half) and $b_{12}$,
the entire sample. We also show $\sqrt{b_1b_2}$, which is consistent with  
$b_{12}$, as expected.
A higher mass cut eliminates more
low bias galaxies and thus increases the bias overall. The lower mass
threshold of $M=1.2\times10^{13}\,M_\odot/h$ reproduces the $b_{12}=1.86$
found by \cite{Gaztanaga2015} for the CMASS sample. The higher mass threshold $M=2.4\times10^{13}\,M_\odot/h$ reproduces the $b_{12}=2.12$ measured by \cite{Alam2016Measurement} for the full CMASS sample (his Fig.~5).
In the top panel of Fig.~\ref{fig:bias_mass}, no extra scatter was added to the
masses (i.e. $\sigma=0$ in Eqn.~\ref{scatter}). The difference
between $b_1$ and $b_2$ is therefore maximised.

\cite{bonvin2014} and  \cite{Gaztanaga2015},
showed that in linear theory the asymmetry of
cross-correlation functions 
(quantified by the dipole, or $z^\text{shell}$)
is proportional to the bias difference $b_1-b_2$. In order to 
match the bias difference of an observed sample, we vary $\sigma$.
In the bottom panel of Fig.~\ref{fig:bias_mass}, we show how 
 $\sigma$ affects the values of $b_1$ and $b_2$. When $\sigma$ is 
large, the larger scatter between $M$ and $M_{\rm obs}$ means that the
two halves overlap much more and consequently have smaller $b_1-b_2$.

\section{The influence of structure on galaxy and halo scales}
\label{sec:halos}
Although there is large-scale structure in the gravitational potential 
(e.g., Fig. \ref{fig:sliceplot}) which contributes to the
asymmetry of clustering signal, the light from galaxies must also exit the
potential well of the subhalo, and this must also be accounted for.
We find that the signal (e.g., from $z^\text{shell}$)
 is in fact very sensitive to the
gravitational potential on small scales, and we explore this
here.
\subsection{Gravitational potentials of subhalos}
The simulated data includes 
gravitational potentials for each particle in each subhalo.
One therefore needs to decide given this information what the
inferred gravitational potential of a galaxy residing in the subhalo would
be. In our previous results in this paper we have 
taken the mean
potential of all particles belonging to a subhalo as the potential of
the galaxy. As luminous galaxies will tend to lie in the densest
part of a subhalo, and therefore trace these regions, then
averaging over the subhalo instead will 
clearly suppress the signal of the gravitational
redshift effect. Another choice is to use the potential of the most
bound particle inside each subhalo. As can be seen from the contours
of the top left panel in Fig.~\ref{fig:sliceplot}, the bound particles
around the centre have significantly higher gravitational 
redshifts.

 Additional sources of difference between the potentials of galaxies and
subhalos could be the influence of galaxy formation
physics (gas cooling, star formation), as well as structure on the scales
of galaxies which we do not resolve in our simulation.
We explore this in more detail with hydrodynamic simulations
in future work (\citealt{Zhu2016Hydro}, in preparation), but for now we can assess
the potential differences using an analytic galaxy model.
We use the result for the gravitational
redshift profile for a galaxy with the approximation
due to \cite{cappi1995} (his Eqn. 7): 
\begin{equation}\label{eq:cappi}
\begin{split}
V(R) & \simeq \frac {GM} {cR_0}\left[1+\ln\left(\frac {R_0} R\right)\right] \\
& \simeq 10^{-5}\sigma_{v}^2\left[1+\ln\left(\frac {R_0} R\right)\right],
\end{split}
\end{equation}
where $R_0=3R_\mathrm{e}$, $R_\mathrm{e}$ is the half-mass radius and $\sigma_v$
is the galaxy velocity dispersion which is correlated with
galaxy mass. As is pointed out
by \cite{cappi1995}, this is a good approximation to
 the projected gravitational
potential profile within the range $0.05R_\mathrm{e}<R<3R_\mathrm{e}$.

So far in our analysis of the simulations we have been using the 
distribution of dark matter particles. We expect the stellar mass
to be more concentrated, however, so that the effective radius
of stars will be different to the dark matter. In order to 
take this into account, we use results from hydrodynamical simulations
to estimate how the stars
 contribute to the potentials via the analytic
potential profile. We randomly choose 20 subhalos with $M_\mathrm{sub}
> 3\times10^{13} M_\odot/h$ in the MBII hydrodynamic simulations
(\citealt{khandai2015}; \citealt{Zhu2016Hydro}) and
compute the half-mass radius of the star $R_\mathrm{e,star}$ and
darkmatter $R_\mathrm{e,dm}$ particles respectively. We use these
results to estimate the increment of
the redshift of galaxies from star particles on
a galaxy by galaxy basis as follows: 
\begin{equation}
\begin{split}
z_\mathrm{galaxy} & = V(R_\mathrm{e,star}) - V(R_\mathrm{e,dm}) \\
&\simeq 10^{-5}\sigma_{v}^2\ln\left(\frac {R_\mathrm{e,dm}}{R_\mathrm{e,star}}\right),
\end{split}
\end{equation}
where the galaxy velocity dispersions ($\sigma_v$) are taken from our $N$-body subhalo catalogues.

%For simplicity, we choose $R$ to be $R_e$ and compute the potential difference to %the center (his Eqn. 5) on a galaxy by galaxy basis. The galaxy velocity dispersions %($\sigma_v$) are taken from the subhalo catalogues.. 

\begin{figure}
  \centering
    \includegraphics[width=0.5\textwidth]{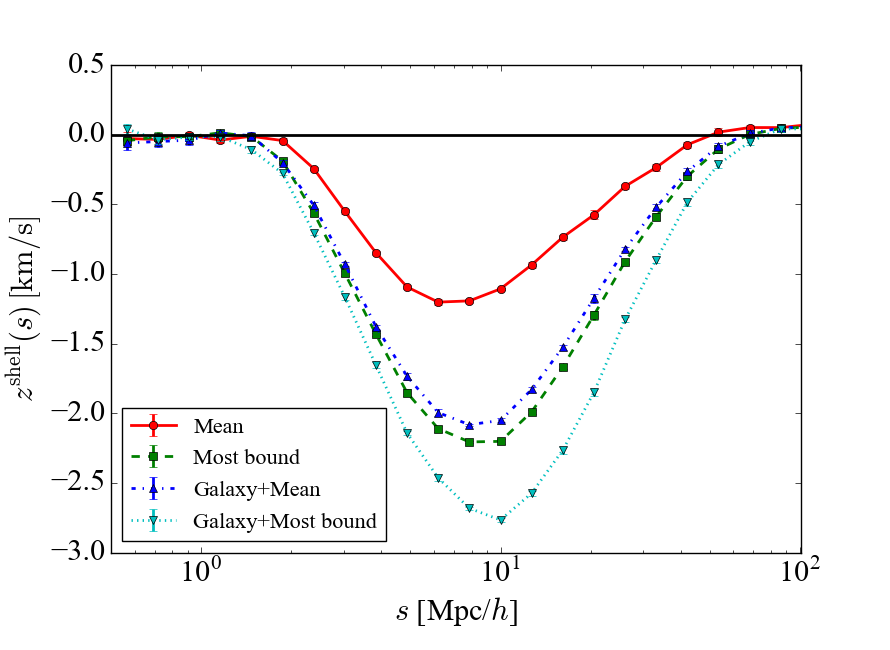}
    \caption{The clustering
asymmetry $z^\text{shell}$ measured from simulations using
different prescriptions for the galaxy potential on subhalo scales. Only
      gravitational redshift effects are included. The circles with
 solid lines 		  (red) show the effect of taking 
the  galaxy potential to be the mean potential of all the
      subhalo particles. The squares with the dashed line (green) show the galaxy potential as the potential of the most bound subhalo particle. The triangles with the dash dotted line (blue) and the upside down triangles with the dotted line
      (cyan) include the effect of
 internal galaxy structure using Eqn.~\ref{eq:cappi}. The error bars are
      the errors on the mean of 8 simulation realizations.}
     \label{fig:galaxy_lss}
\end{figure}

In Fig.~\ref{fig:galaxy_lss} we show how the different 
means of modelling the galaxy potential from the subhalos 
affect the clustering asymmetry due to $z^\text{shell}$. We only include the
effect of gravitational redshifts here, although a full accounting
should also include relativistic effects due to beaming  and transverse Doppler
shifts of stars for example.
 We can see by adding galaxy
structure information onto the fiducial simulation (which only
includes large-scale structure), 
the amplitude  of $z^\text{shell}$ can increase dramatically.
In fact for the example  plotted in Fig.~\ref{fig:galaxy_lss},
most of the asymmetry is due to structure on 
galactic scales, even when measured using the cross-correlation
function on 10 Mpc/$h$ scales. The fact that there is this dramatic difference
underscores the need for a better understanding of the small-scale
contributors to relativistic clustering distortions. The largest 
uncertainty in our current predictions is therefore due to this
small-scale structure, and this should be borne in mind when comparing
to observations. We  note that although the mean and most bound potentials 
have a significantly different amplitude on their own, when including a 
 galaxy component the difference between mean and most bound potential 
becomes small.
\begin{figure}
  \centering
    \includegraphics[width=0.5\textwidth]{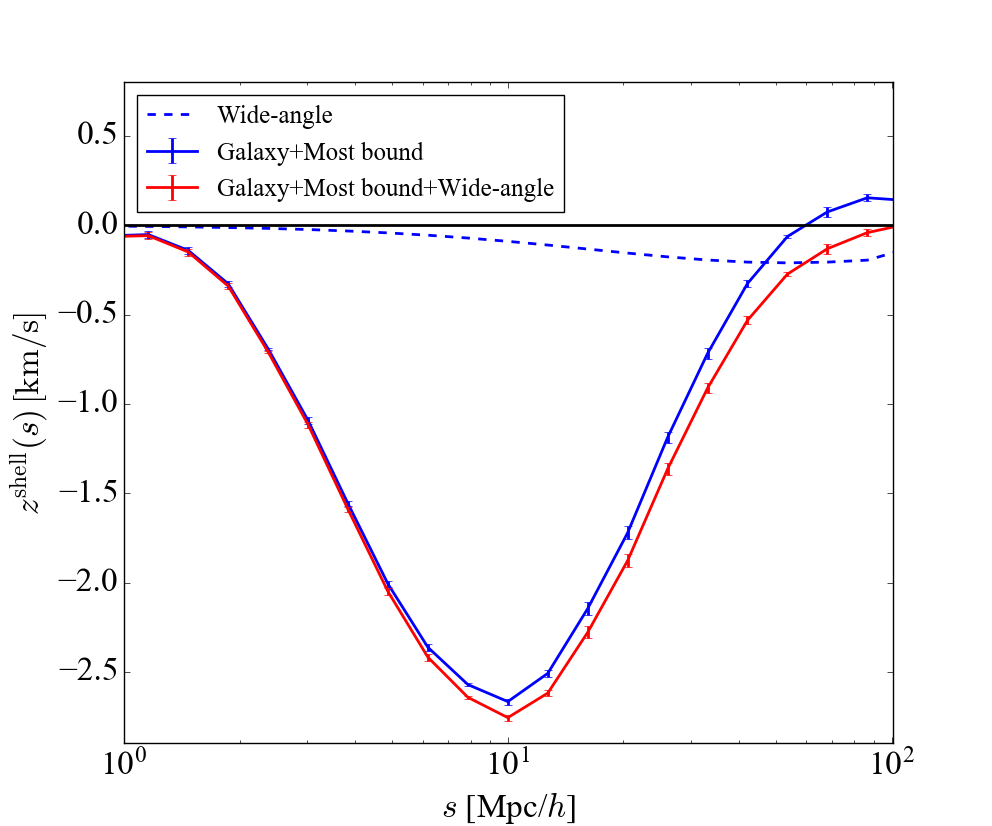}
    \caption{The clustering asymmetry $z^\mathrm{shell}$ from simulations including all the components discussed above. The dashed blue line represents the wide-angle effects computed in \protect\cite{Giusarma2016PT}. The solid blue line includes the five effects discussed above (computed using a boost factor of 50) but without the wide-angle effects. The solid red line adds the wide-angle effects on top of the solid blue line, and this is used as the model fit in our companion paper \protect\cite{Alam2016Measurement}. The error bars are the errors on the mean of 8 simulation realizations.}
    \label{fig:est_shadab}
\end{figure}

We have chosen the multiplication factor appropriate for making low noise
 predictions
of the relativistic asymmetry ( see Sec.~\ref{sec:shell_dipole} for 
discussion of this factor). In Figure 
Fig.~\ref{fig:est_shadab} we show the component due to wide-angle effects as well as the associated $z^\text{shell}$ results from
the simulation, which we use as our model fit in the  measurement
of clustering statistics in the BOSS/CMASS galaxy 
sample \cite{Alam2016Measurement}. We see that the wide-angle effects lead to a $\sim -0.2$ km/s offset in this case, which is small compared with overall amplitude of Gravitational redshift effects.

\subsection{Resolution test}
The resolution of the simulation can affect the shell estimator as
well. If the simulation cannot resolve the subhalos very well, the
signal will also be suppressed. We construct a resolution test for only one realization in a smaller box with side length 125 Mpc/$h$ by
running new simulations of the same cosmological model, but with
higher mass and spatial resolution,
doubling, tripling and quadrupling the number of particles per Mpc/$h$ 
along each
dimension compared to our fiducial simulations. The numbers of particles along each dimensions are 128, 256, 384, 512 so that the resolutions are ``resolution'', ``resolution$\times2^3$'', ``resolution$\times3^3$'' and ``resolution$\times4^3$'' accordingly. We change the mass and force resolution together and thus the softening length becomes 2.5 kpc/$h$.

\begin{figure}
  \centering
    \includegraphics[width=0.5\textwidth]{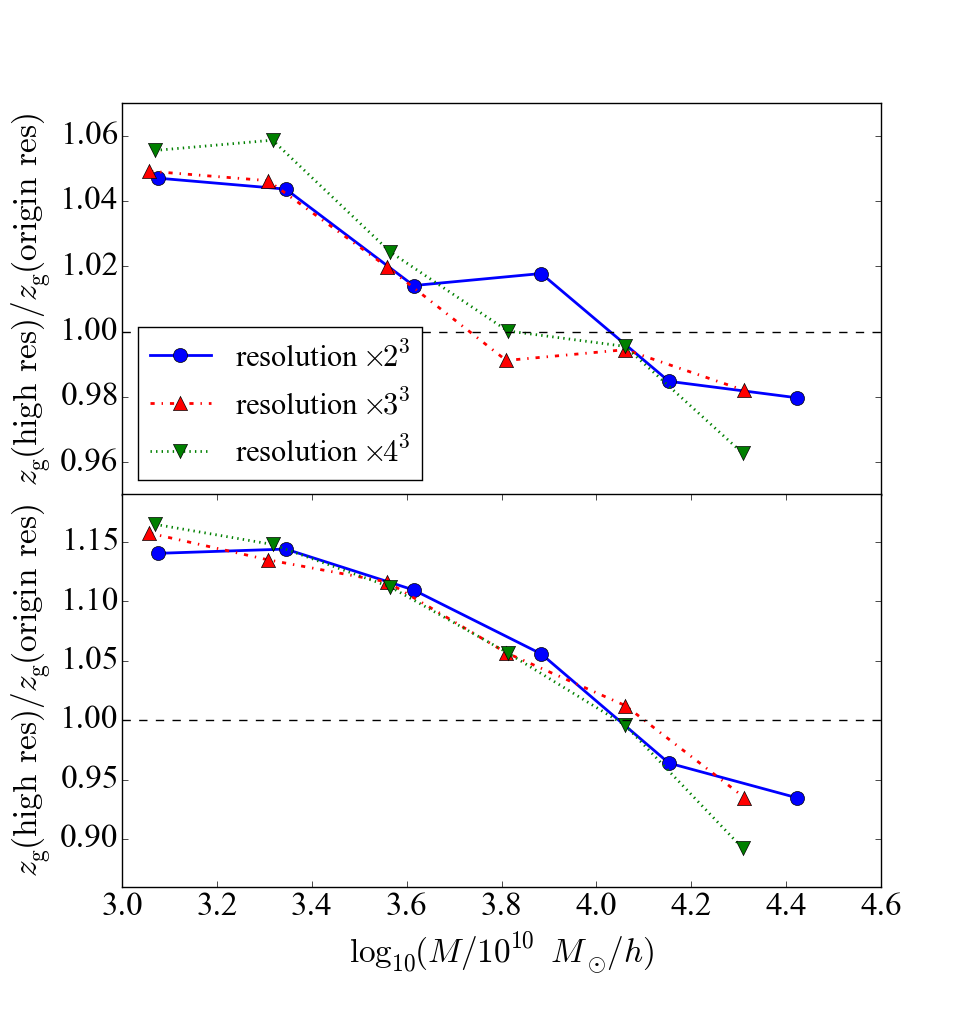}
    \caption{
Resolution test of the subhalo potentials. Each
      panel shows the  mean value of the ratio of 
$z_\mathrm{g}$, the gravitational
redshift measured from a set of halos in 
higher resolution simulations divided by the results for the same halos
(see text for matching) 
in a simulation with our fiducial resolution. The top panel
associates the galaxy potential with the mean
potential of all particles in the subhalo. In the bottom  panel we show
results when the galaxy potential is associated with the potential 
of the most bound subhalo particle.}
     \label{fig:res}
\end{figure}

We run the simulations using the same initial random phases and 
the same box sizes, so that we are able to match subhalos in the 
different resolution runs according to their positions. We then
compare the $z_\mathrm{g}$ (gravitational redshifts)  of the same subhalos
between different resolutions.
The results for the mean ratio of $z_\mathrm{g}$ as a function
of subhalo mass for the different runs. are shown in 
 Fig.~\ref{fig:res}. The top panel was computed using
the mean potential of subhalo particles and
the bottom panel from the most bound subhalo. As expected, the difference
due to resolution for the latter is greater. We can also see that the
results have effectively converged with a resolution which 
is twice as good as our fiducial resolution. The 
maximum differences
in $z_\mathrm{g}$ between our fiducial 
simulation and the converged run are between 5\% and 15\%. Given the
much larger uncertainties in structure on galactic scales discussed
above, this is acceptable.

\section{Summary and discussion}
\label{sec:discussion}
\subsection{Summary}
Using $N$-body simulations, we have modeled different
relativistic effects which cause small asymmetries in the
clustering of galaxies. These effects, such as the
gravitational redshift have previously been observed in
galaxy clusters (e.g., \citealt{wojtak2011}). We have extended
the study of other effects such as the transverse Doppler effect
and relativistic beaming modeled in clusters by 
\cite{zhao2013} and K13 to the case of large scale structure.
We have focussed on non-linear scales $\sim 10$ Mpc/$h$ and below, where
perturbation theory results (e.g., \citealt{yoo2009}; \citealt{yoo2012}; \citealt{bonvin2014}) are
less likely to be accurate, and followed the quasi-Newtonian
approach taken by C13 for gravitational redshifts. 
We have made predictions for the
distortions in the galaxy cross-correlation function of galaxies with 
different mass and bias, and quantified the
line-of-sight asymmetries. 
Our conclusions are as follows:
\begin{enumerate}
\item The centroid of the galaxy high mass versus low mass
cross-correlation function is shifted in the line-of-sight 
direction
by a maximum of around 2-3 km/s on scales $\sim$ 10 Mpc/$h$. 
This is smaller than the 
redshift profile shift seen in galaxy clusters (e.g., \citealt{wojtak2011})
by around an order of magnitude.

\item All five relativistic effects we include 
 (the gravitational redshift effect, the
  TD, LC, SRB and LDP effects) lead to asymmetries in
  the cross-correlation functions, but with differences in magnitude,
sign and variation with scale. The LC and TD effects cause a
positive shift ($z^{\rm shell}$ positive) on scales $s \sim 10 $
Mpc/$h$. The asymmetry signal is
 however  dominated  by the gravitational redshift effect on scales 
  $s \sim   10$ Mpc/$h$, and has 
 opposite sign. At larger scales 
 the effect of luminosity distance perturbations and
the effect of special relativistic beaming on galaxy selection 
becomes more important, which again leads to an overall positive
shift in $z^{\rm shell}$. Finally, wide-angle effects are added, which depend on the galaxy bias differences. These lead to an overall small blueshift or redshift offset to the signal on large-scales.
\item We compare two estimators of clustering 
asymmetry-- the shell estimator and the dipole.
  We find that the shell estimator is mostly independent of the 
  multiplication factor used in simulations to increase the signal
to noise, while the dipole moment is not. Because of this we
focus on the shell estimator, and adjust the multiplication
factor so that the statistical and systematic errors from our simulation
predictions are approximately equal (and below 5\%).
\item 
We find that approximately half or more of
 the observable asymmetry signal is likely
caused by potential gradients from  structure on galaxy
scales  instead of the large-scale information.
\end{enumerate}

\subsection{Discussion}

Some of the uncertainties in our predictions, can be 
traced to small-scale physics and the 
uncertainties in the subhalo gravitational potentials. For example, we have
experimented  using different ways to map the potentials
of particles in subhalos to the potentials of galaxies -- the mean and 
the potential of the most bound particle, finding 
approximately twice the signal using the potential of the most bound particle.
We have also investigated the influence of structure on galactic scales, 
adding the contribution of a galaxy potential profile, and find that this
can strongly affect the shell 
estimator of asymmetry. 
The galaxy structure can therefore be thought to be the main uncertainty,
(which we have seen can vary the signal by a factor of two).
 It will therefore require further work before we are able to make
predictions for redshift asymmetries expected in the LCDM model at the
few tens of percent accuracy expected to be relevant for 
testing modified gravity theories
 (e.g., \citealt{wojtak2011}; \citealt{gronke2014}).
Not only dark matter but also baryons 
will contribute to the galaxy structure (for example,
through the processes of cooling and feedback). 
In this paper, we have used purely $N$-body simulations but we explore 
hydrodynamic effects using hydrodynamic simulations in a companion paper 
(\citealt{Zhu2016Hydro}, in preparation). 

On the observational side, there
are many promising current and future datasets which 
can be used to measure these relativistic clustering
asymmetries. We have used our work on simulations to develop
a model relevant to the CMASS galaxy sample
from the SDSS BOSS redshift survey. Our observational measurements and
comparison with this model are presented in 
a companion paper (\citealt{Alam2016Measurement}). In the future, surveys such as 
 eBOSS (\citealt{eboss}) and 
DESI (\citealt{desi})
will be used to push the measurement of gravitational redshift
and other asymmetries of clustering forward. On 
small scales, the potential profiles inside galaxies can 
in principle be measured by stacking integral field
spectra of large numbers of galaxies. This
could allow tests of relativistic effects including gravitational
redshifts and beaming of stars on galactic scales, which we have shown
our knowledge is uncertain. The SDSS MaNGA (\citealt{bundy2014}) survey of 
large numbers of galaxies will be a useful dataset in this regard
and we have started investigating this area.

On the theory side, although we have used $N$-body simulations 
here, it should be possible to develop a halo model (\citealt{cooray2002}) for all relativistic clustering effects (TD, LC,
SRB and LDP). This would be along the lines of what was done  by C13, 
for example to make predictions for the 
gravitational redshift. Such a semi-numerical model might enable a range
of models to be covered more rapidly, although testing with
$N$-body simulations will be needed.  Alternative theories to GR (modified gravity, \citealt{clifton2012}) have taken on
more prominence during the
past decade. Modified gravity models may make different predictions
for the clustering asymmetries and they could also be simulated. 

\begin{figure}
  \centering
    \includegraphics[width=0.5\textwidth]{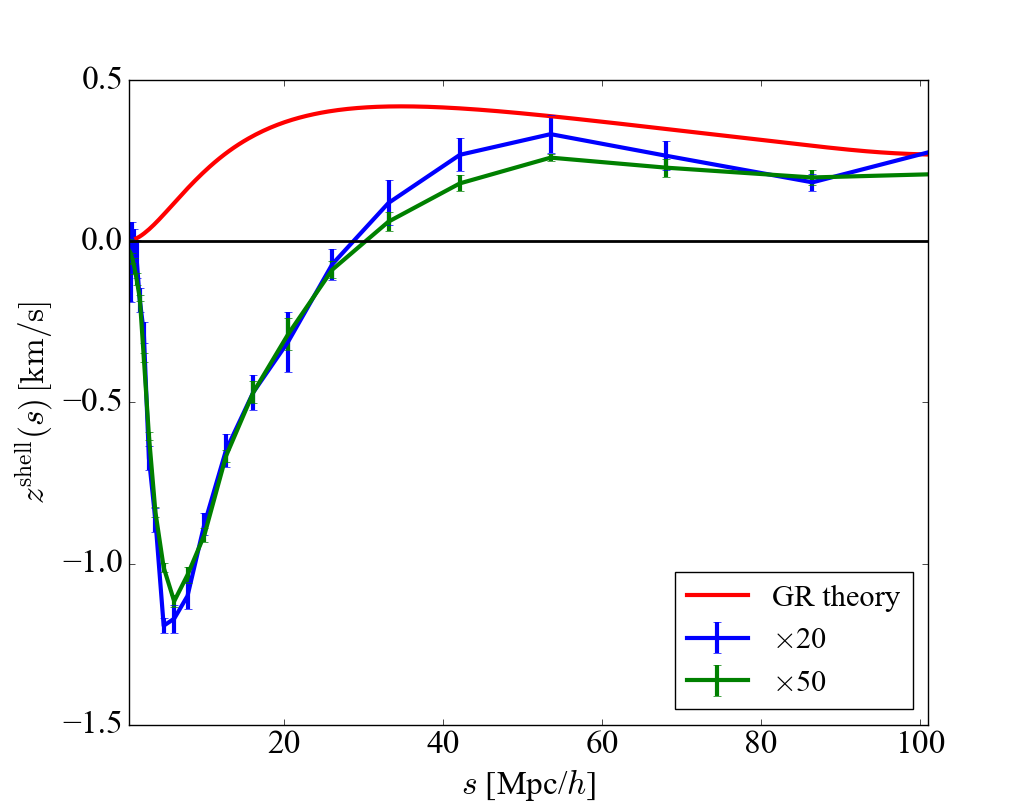}
    \caption{Comparison between our shell estimator with different boost factors (blue, green) and the linear perturbation theory (red). The biases ($b_1=2.57$, $b_2=1.91$) are chosen at $M_\mathrm{cut} = 3\times10^{10}\,M_\odot/h$ and $\sigma_M=0$. Though they disagree on small scales but converge at sufficient large scales.}
     \label{fig:LT}
\end{figure}

% Compare with linear theory
It is useful to compare our results to the GR perturbation
theory predictions in the literature (e.g., \cite{bonvin2014}), as
we should expect a level of agreement on linear scales. We have done
this in Fig.~\ref{fig:LT}, where we use a linear $x$-axis 
to better show the range of scales where predictions can be compared.
To compute the linear pertubation theory curves, we have followed the
  methodology of \cite{Giusarma2016PT}  (based on \cite{bonvin2014}),
  applying the relevant galaxy biases for our samples. We can see that 
 large scales
  ($> 30$ Mpc/$h$, our the shell estimator from the simulations
is quite close to that from linear
  perturbation theory, and has the correct sign.
They are not expected to be identical because
 proper modelling of the special relativistic beaming effect 
requires using the colour-magnitude selection appropriate for BOSS galaxies,
 which we have done in the simulation, but which is not accounted for explicitly in the linear perturbation theory result.

In this paper we have used a quasi-Newtonian approach, with $N$-body simulations
to model non-linear effects. The reasoning behind this is that 
the largest uncertainties are due to these non-linearities (including for
example structure on galactic scales), and the signal to noise of 
future measurements from large-scale structure and current measurements
from clusters is dominated by non-linear scales. The
fully General Relativistic perturbation theory (PT) approach has been
used successfully by e.g., \cite{yoo2009}, \cite{bonvin2014} to make predictions
for relativistic clustering on large, linear scales. Comparison of
the two approaches is needed to find areas of overlap and explore the
consistency of predictions (see for example, \citealt{Giusarma2016PT}, in preparation).

\section*{Acknowledgments}
This work was supported by NSF Award AST-1412966. SA is also supported by the European Research Council through the COSFORM Research Grant (\#670193). We would like to thank Volker Springel for the use of the \textsc{P-Gadget3} code. In addition, our $N$-body simulations used resources of the National Energy Research Scientific Computing Center (NERSC), a DOE Office of Science User Facility supported by the Office of Science of the U.S. Department of Energy under Contract No. DE-AC02-05CH11231.
H. Z. would also like to thank Alexie Leauthaud, Kevin Bundy, David V. Stark and Pengjie Zhang for useful discussions. We thank the referee for pointing out
a number of issues which needed to be addressed.

%%%%% Bibliography %%%%%%%%%%%%%%%%%%%%%%%%%%%%%%%%%%%%%%%%%%%%%%%%%%%%%%%%%%%%

\bibliography{Master_Hongyu.bib}
\bibliographystyle{mnras}

\label{lastpage}

\end{document}